\documentclass[sigconf]{acmart}
\AtBeginDocument{%
  }

\setcopyright{acmlicensed}
\copyrightyear{2018}
\acmYear{2018}
\acmDOI{XXXXXXX.XXXXXXX}
\acmConference[Under Review]{}{}{}
\acmISBN{978-1-4503-XXXX-X/2018/06}

\usepackage{xcolor}
\usepackage{tcolorbox}
\usepackage{bm}
\usepackage{booktabs} 
\usepackage{multirow}
\usepackage{graphicx}
\usepackage{subcaption}
\usepackage{float} 
\usepackage{hyperref}
\usepackage{xcolor} 
\definecolor{mypink}{RGB}{255, 105, 180} 

\theoremstyle{plain}

\theoremstyle{definition}

\theoremstyle{remark}

\begin{document}

\title{Taiji: Pareto Optimal Policy Optimization with Semantics-IDs Trade-off for Industrial LLM-Enhanced Recommendation}
\author{Yuecheng Li}
\affiliation{%
  \institution{Kuaishou Technology}
  \city{Beijing}
  \country{China}
}
\email{liych168168@gmail.com}


\author{Zeyu Song}
\affiliation{%
  \institution{Kuaishou Technology}
  \city{Beijing}
  \country{China}
}
\email{songzyzzz@gmail.com}


\author{Jing Yao}
\affiliation{%
  \institution{Kuaishou Technology}
  \city{Beijing}
  \country{China}
}
\email{airanabetter07@gmail.com}

\author{Chi Lu}
\affiliation{%
  \institution{Kuaishou Technology}
  \city{Beijing}
  \country{China}
}
\email{looc0727@gmail.com}

\author{Peng Jiang}
\affiliation{%
  \institution{Kuaishou Technology}
  \city{Beijing}
  \country{China}
}
\email{jp2006@139.com}

\author{Kun Gai}
\affiliation{%
  \institution{Unaffiliated}
  \city{Beijing}
  \country{China}
}
\email{gai.kun@qq.com}

\begin{abstract}
  Scaling recommender systems via large language models (LLMs) has become a prominent trend in the industry. However, aligning the LLM's semantic space with the recommender's ID space via post-training (e.g., SFT and RL) remains challenging. Existing LLM4Rec paradigms are bottlenecked by two main issues: \textit{(1) the difficulty of measuring and improving chain-of-thought (CoT) quality in open-domain recommendation during SFT}, and \textit{(2) the neglect of the trade-off between LLM semantic rewards and recommendation preference rewards during RL alignment}. Inspired by these challenges, we present \textbf{Taiji}, a novel LLM-as-Enhancer framework designed for industrial recommender systems. To overcome the SFT bottleneck, we utilize reverse-engineered reasoning and open-ended rejection sampling to generate high-quality, domain-specific CoT data. To resolve the RL alignment issue, we propose \textbf{P}areto \textbf{O}ptimal \textbf{P}olicy \textbf{O}ptimization (\textbf{POPO}), which adaptively adjusts cross-domain reward weights. Theoretically, it achieves an optimal trade-off between the semantic world knowledge of LLMs and the collaborative ID features representing online user preferences. Extensive offline evaluations and online A/B tests validate the effectiveness of Taiji. Deployed on Kuaishou's advertising platform since May 2026, Taiji currently serves over 400 million users daily, yielding significant commercial revenue and demonstrating its robust scalability in web-scale environments.
\end{abstract}




\maketitle

\section{Introduction}

With the deep integration of LLMs and recommender systems, the field of Large Language Models for Recommendation (LLM4Rec) has progressively evolved into three primary paradigms: \textit{Generative Recommendation} \cite{rajput2023tiger, zhai2024hstu, deng2025onerec}, \textit{Ranking Model Scaling} \cite{zhu2025rankmixer, jiang2026tokenmixer, zhang2026onetrans}, and \textit{LLM-as-Enhancer} \cite{xi2024KAR, gu2025r4ec, lin2025Rec-R1, chen2025deeper, gao2025langptune}. Through distinct technical pathways, these three paradigms aim to unleash the generalization and scaling capabilities of LLMs within large-scale recommender systems. Among them, the \textit{LLM-as-Enhancer} paradigm has achieved the most widespread adoption in industrial applications, primarily due to its architectural decoupling from online serving models and relatively controllable costs \cite{xi2024KAR, xia2025hit-lbm, xia2025trackrec, li2026recgoat}. Specifically, this paradigm leverages frozen pre-trained or post-trained LLMs to generate semantic representations of user profiles or item content, thereby augmenting the input features of downstream recommendation backbones.

To bridge the semantic gap between LLMs and recommendation tasks, existing \textit{LLM-as-Enhancer} approaches can be broadly categorized into three progressive methodologies: Direct Inference, Domain Fine-tuning, and Reinforcement Alignment. 
Early works rely on Direct Inference, where models like KAR \cite{xi2024KAR} and HiT-LBM \cite{xia2025hit-lbm} utilize prompting strategies or tree-search mechanisms to extract user preferences and item factual knowledge from LLMs, converting them into augmented vectors compatible with arbitrary recommendation models.
To further enhance domain-specific reasoning, subsequent methods adopt Domain Fine-tuning. Frameworks such as $R^4$ec \cite{gu2025r4ec}, and TrackRec \cite{xia2025trackrec} employ supervised fine-tuning (SFT) and iterative refinement mechanisms to generate reliable, recommendation-specific Chain-of-Thought (RecCoT) data, mitigating the reasoning hallucinations of LLMs.
More recently, to strictly align LLM outputs with downstream recommendation objectives, \textit{Reinforcement Alignment} has gained significant traction. Approaches including DEEPER \cite{chen2025deeper}, RecLM \cite{jiang2025reclm}, Rec-R1 \cite{lin2025Rec-R1}, LangPTune \cite{gao2025langptune}, and RecGPT-v2 \cite{yi2025recgpt} leverage reinforcement learning algorithms (e.g., PPO, DPO, GRPO) to directly optimize the LLM's generation policy using recommendation metrics (e.g., NDCG, Recall) or user feedback  (e.g., click-through rate, user retention) as reward signals. This paradigm effectively mitigates the exposure bias inherent in supervised fine-tuning by directly optimizing real-world recommendation objectives, thereby aligning model outputs more closely with business goals.

However, existing post-training paradigms still face significant limitations across different stages:

\begin{enumerate}
    \item \textbf{In the SFT stage}: Previous works often over-rely on the CoT generation capabilities of powerful teacher LLMs \cite{xi2024KAR, jiang2025reclm} or heuristic CoT refinement experiences \cite{gu2025r4ec}. Alternatively, they evaluate CoT quality solely based on the correctness of the final answer \cite{xia2025trackrec}. \textit{Due to the open-domain nature of recommendation tasks, there remains a lack of reasonable and systematic metrics to accurately measure the quality of recommendation-specific CoT.}
    \item \textbf{In the RL stage}: Although current methods utilize LLM semantic rewards \cite{jiang2025reclm, lin2025Rec-R1} and recommendation feedback rewards \cite{lin2025Rec-R1, gao2025langptune} for preference optimization, they fail to consider the dynamic balance between these two types of heterogeneous information. \textit{For the LLM-as-Enhancer paradigm, effectively aligning and balancing the world knowledge semantics of LLMs with the online user preferences of recommender systems is a critical issue}. Existing approaches focus merely on alignment, lacking a profound trade-off mechanism. 
\end{enumerate}

To tackle these two critical challenges, we propose \textbf{Taiji} (named after the Taiji/Yin–Yang diagram to symbolize the dynamic unity and mutual reinforcement between LLMs and recommender systems), an industrial-scale \textit{LLM-as-Enhancer} comprising four key modules: \textit{Data Construction}, \textit{Reasoning Activation}, \textit{LLM-Recommendation Collaboration}, and \textit{Online Ranking}. First, we collect a large-scale dataset from real-world online video streaming logs on the Kuaishou platform, encompassing user profiles and their recent behavioral sequences. Subsequently, we introduce \textbf{Reverse-Engineered User Preference Reasoning (EUPR)}. By utilizing ground-truth user-item collaborative relations as prompts, we distill high-quality reasoning CoT data from the advanced QwQ-32B \cite{qwq-32b-preview}. Next, we propose \textbf{Open-Ended Rejection Sampling Fine-Tuning (ORFT)}, which filters out low-quality CoT samples based on the Perplexity (PPL) metric and performs SFT on the DeepSeek-R1-7B \cite{guo2025deepseek-r1}. To effectively balance the LLM semantic rewards and recommendation preference rewards, we propose \textbf{Pareto Optimality Policy Optimization (POPO)}, enabling a comprehensive exploration of the Pareto fronts within the cross-domain space. Finally, the outputs from the RL-aligned LLM are encoded to construct quantized sparse features and retrieved cross-user sequences,respectively, which are subsequently incorporated into the online advertising ranking model.

Overall, the primary contributions of this paper are summarized as follows:

\begin{itemize}
    \item We present Taiji, a novel framework for industrial LLM-enhanced recommendation, which addresses two core limitations during the post-training stage of the conventional LLM-as-Enhancer pipeline. Specifically, in the SFT stage, we integrate EUPR and ORFT to asymptotically enhance the quality of recommendation-specific CoT. Furthermore, in the RL stage, we introduce POPO, which dynamically adapts the weights of LLM semantic rewards and recommendation preference rewards, achieving a theoretically guaranteed Pareto optimal trade-off between heterogeneous information.
    \item We conduct extensive offline experiments and ablation studies, rigorously validating the effectiveness of each module within Taiji. Furthermore, online A/B testing demonstrates that Taiji yields a \textbf{2.83\%} overall ADVV improvement for Advertiser Value and drives a \textbf{3.30\%} increase in overall Revenue for Kuaishou's Advertising platform.
    \item Taiji has been fully deployed in the production environment since May 2026, stably supporting over 400 million daily active users (DAU).

\end{itemize}

\begin{figure*}[t]
\centering
\includegraphics[width=6.5in]{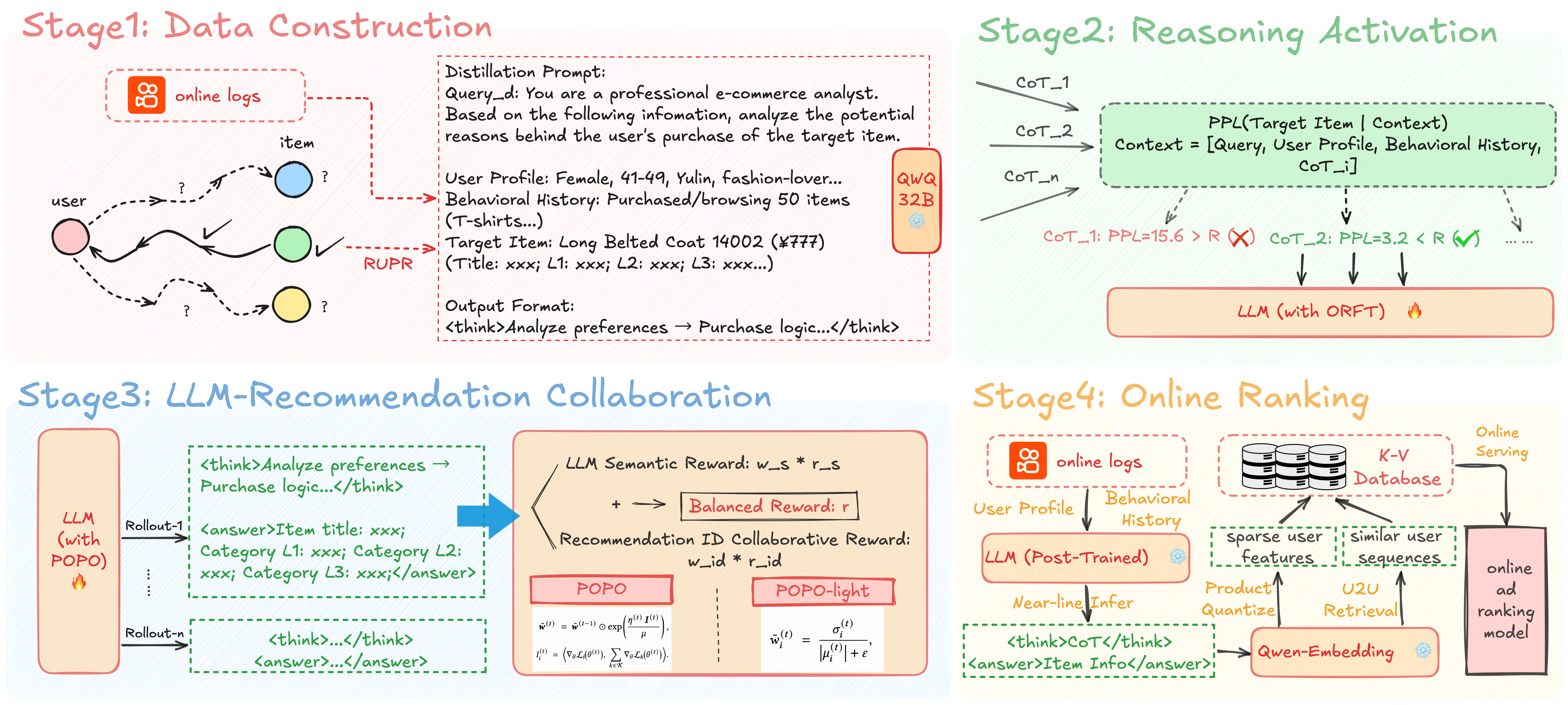}
\caption{The overall framework of Taiji. It consists of four stages: (1) Data Construction. Taiji collects real-world data from the production platform, constructs distillation prompts via reverse engineering, and leverages a teacher LLM to generate user preference CoT reasoning. (2) Reasoning Activation. Taiji filters low-quality CoT samples using PPL-based rejection sampling and performs supervised fine-tuning on a 7B LLM. (3) LLM-Recommendation Collaboration. Taiji proposes POPO to dynamically adjust cross-domain reward weights during the RL process. (4) Online Ranking. Taiji performs near-online inference on live user data using the RL-optimized LLM, generating quantized sparse features and retrieved cross-user sequences as inputs to the online ad ranking model.}

\label{g3}
\end{figure*}

\section{Methodology}

To enhance the quality of CoT reasoning during the SFT stage and optimize the cross-domain information trade-off in the RL stage within the \textit{LLM-as-Enhancer}, while seamlessly adapting to industrial recommenders, we propose a novel framework named \textbf{Taiji}. Specifically, Section \ref{sec:2.1} details the construction of the offline dataset and the generation process of recommendation-specific CoT, which serve as the foundation for the subsequent SFT (Section 2.2) and RL (Section 2.3) training phases. Finally, Section 2.4 presents our enhancement strategies for the online advertising recommendation model. The overall pipeline is illustrated in Figure \ref{g3}.


\subsection{Data Construction: Reverse-Engineered User Preference Reasoning}
\label{sec:2.1}

To align LLMs with recommender systems, we sample extensive user data from the real-world online logs of a short-video platform for the post-training of the LLM. Furthermore, unlike mathematical reasoning or code generation, LLM-based recommendation is inherently an open-ended generation task. Consequently, when distilling CoT data from a teacher model, it is challenging to precisely verify the accuracy of the generated reasoning trajectories and the final answers. Therefore, we employ Reverse-Engineered User Preference Reasoning (RUPR) to generate preliminary, relatively accurate CoT while ensuring the ground-truth validity of the answers.


\subsubsection{Data Collection}
\label{sec:2.1.1}

To construct high-quality datasets for instruction fine-tuning and reinforcement learning, we collect large-scale user profiles and behavioral sequences from real-world online logs on Kuaishou, and convert them into LLM-friendly natural language texts. Specifically, our data collection consists of two key components:
\begin{itemize}
    \item \textbf{Multimodal and Multidimensional User Profile}. We integrate multiple data tables to extract comprehensive user features. These encompass basic demographics (e.g., gender, age, city tier, marital status, education), device and lifestyle attributes (e.g., phone model, residential consumption level), and multimodal interaction preferences on the short-video platform (e.g., active apps, search queries, video engagements like saves/likes/comments, live-streaming interactions, and historical e-commerce/ad behaviors). These fine-grained features are serialized into structured natural language descriptions to provide rich personalized contexts for LLMs.
    \item \textbf{Fine-Grained Historical Behavioral Sequence}. We collect users' recent ad interaction logs, including item views and purchases. To ensure timeliness, we retain the 50 most recent interactions in reverse chronological order. For each interaction, detailed item meta-information (e.g., title, multi-level categories, price) is extracted and templated into text sequences (e.g., \textit{"Item title: ...; Category L1: ...; Category L2: ...; Category L3: ..."}). This enables the LLM to accurately capture users' evolving interest and purchasing logic.
\end{itemize}

Through this collection and text serialization process, we transform conventional tabular recommendation features into semantically rich prompts, laying a solid foundation for subsequent CoT generation, model fine-tuning, and RL-based alignment.

\subsubsection{CoT Generation}
\label{sec:2.1.2}
For open-ended generation tasks such as LLM-based recommendation, inspired by \cite{wang2025reer, RE-AdaptIR}, we propose Reverse-Engineered User Preference Reasoning (RUPR) to generate reliable initial CoTs. Specifically, based on the collected online data, we construct \textit{distillation prompts} for RUPR, as depicted in Stage 1 of Figure \ref{g3}. Meanwhile, we utilize QWQ-32B \cite{qwq-32b-preview} to perform reverse preference reasoning by conditioning 
on the user's ground-truth next purchased item.

\subsection{Reasoning Activation: Open-Ended Rejection Sampling Fine-Tuning}
\label{sec:2.2}
In this section, we employ open-ended rejection sampling to further filter and obtain high-quality CoT data. Subsequently, we perform supervised fine-tuning (SFT) to activate the reasoning capabilities of the smaller model, thereby enhancing the LLM's comprehension of user preferences and its proficiency in item recommendation.

\subsubsection{CoT Refinement}
To further improve the quality of recommendation-oriented CoTs, we use the \textbf{perplexity (PPL)} of the user’s ground-truth next purchased item \(y\) (enclosed by the \colorbox{gray!10}{\texttt{<answer>...</answer>}} tag) as a proxy for the quality of a given reasoning CoT (enclosed by the \colorbox{gray!10}{\texttt{<think>...</think>}} tag). The computation is defined as:
\begin{equation}
\begin{aligned}
\mathrm{PPL} &= \exp\!\left(-\frac{1}{T}\,\log\_likelihood\right), \\
\log\_likelihood &= \sum_{t=1}^{T}\log P\!\left(y_t \mid \text{context},\, y_1,\ldots,y_{t-1}\right).
\end{aligned}
\end{equation}
where context=[\textit{query}, \textit{user profile}, \textit{behavioral history}, \textit{CoT}], $y=[y_1,y_2,\ldots,y_T]$ represents the ground-truth answer sequence, and T is the number of tokens. A lower PPL indicates that the CoT assigns higher probability mass to the ground-truth item, suggesting a more reliable reasoning trajectory.

\begin{figure}[t]
\centering
\includegraphics[width=3.3in]{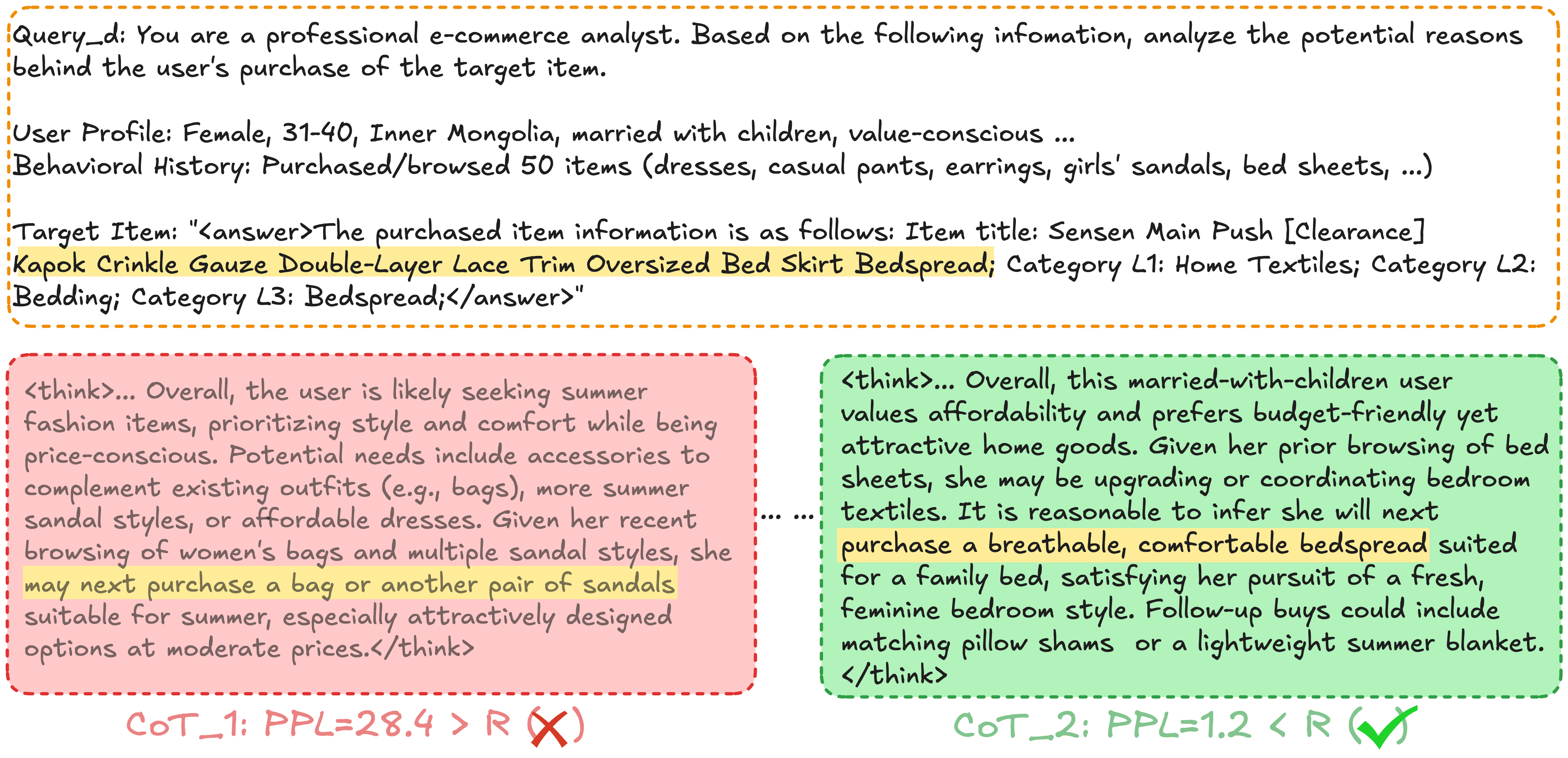}
\caption{Illustration of Open-Ended Rejection Sampling for CoT Refinement. Given the same user profile, behavioral history and target item, two candidate CoTs are generated by QwQ-32B. $CoT_1$ (left, red box) incorrectly infers summer fashion needs and yields high perplexity (PPL = 28.4 > $R$), indicating poor alignment with the ground-truth bedspread purchase, and is thus rejected. In contrast, $CoT_2$ (right, green box) accurately reasons about the user's home textile upgrade intent, achieving low perplexity (PPL = 1.2 < $R$) and being retained for fine-tuning.}
\label{g2}
\end{figure}

Concretely, we leverage the advanced reasoning model QwQ-32B \cite{qwq-32b-preview} to generate \(k=3\) candidate CoTs for each user-specific distillation prompt, and retain those satisfying \(\mathrm{PPL}<R\) as training data for subsequent fine-tuning (yielding 0–3 retained reasoning paths per prompt). Here, $R$ is a preset PPL cutoff threshold, empirically determined as the median (50th percentile) of the PPL distribution computed over a sampled subset of the data. As illustrated in Figure \ref{g2}, when PPL is excessively high, logical inconsistencies emerge between the CoT and the answer, indicating that such training samples are highly likely to degrade the LLM's reasoning performance.

\subsubsection{LLM Fine-Tuning}
To activate the reasoning capabilities of LLMs for recommendation tasks, we perform supervised fine-tuning on DeepSeek-R1-7B \cite{guo2025deepseek-r1} using the refined CoT data. Specifically, the SFT training samples are constructed as follows:

\tcbset{colback=gray!10, colframe=gray!88, boxrule=0.5pt, arc=2pt}

\begin{tcolorbox}[title=SFT Prompt, fonttitle=\bfseries]
\textbf{Input}

 - \textbf{Query}: You are a professional e-commerce analyst. Based on the following information, reason about why the user might purchase the advertised item and predict potential follow-up purchases.

 - \textbf{User Profile}: \texttt{Female, Age: 31-40, Inner Mongolia, Married with children, Value-conscious ...}

- \textbf{Behavioral History}: \texttt{Purchased/browsed 50 items (dresses, casual pants, earrings, girls' sandals, bed sheets, ...)}

\textbf{Output}

\texttt{<think> Refined CoT  </think>}; 
\texttt{<answer> \textbf{Item title}: Sensen Main Push [Clearance] Kapok Crinkle Gauze Double-Layer Lace Trim Oversized Bed Skirt Bedspread; \textbf{Category L1:} Home Textiles; \textbf{Category L2}: Bedding; \textbf{Category L3}: Bedspread;</answer>}
\end{tcolorbox}

\textbf{Training Objective}. The training loss for this fine-tuning process is formulated as a standard next-token prediction objective over the reasoning chain and the final answer:
\begin{equation}
\mathcal{L}_{\text{ORFT}} = -\frac{1}{N} \sum_{i=1}^{N} \sum_{j=1}^{l_i} \log P(t_j \mid q_i, t_{<j}),
\end{equation}
where $q_i$ denotes the input query (including user profile and behavioral history), $t_j$ represents the $j$-th token in the complete output sequence (i.e., both the \texttt{<think>} CoT and \texttt{<answer>} sections), $l_i$ is the total number of tokens in the output, and $N$ is the batch size.

\subsection{LLM-Recommendation Collaboration:  Pareto Optimality Policy Optimization}
\label{sec:2.3}
After the SFT stage, the reasoning capabilities of the LLM in the recommendation domain are initially activated. To further enhance the LLM's generalization ability across diverse user-item interaction patterns, we propose a GRPO-like reinforcement learning process in this section. However, prior work typically applies fixed weights to combine semantic rewards from LLMs and preference-based rewards from recommender systems \cite{zhang2025gpr, zhang2026onemall}, making it difficult to achieve a dynamic equilibrium between the two objectives. Specifically, from an effectiveness perspective, static reward weights struggle to capture the complex, non-convex Pareto front. From an optimization efficiency perspective, the inherent differences in learning difficulties among various reward signals often lead to suboptimal convergence rate. To address these limitations, we propose \textbf{Pareto Optimality Policy Optimization (POPO)}, which employs dynamic reward weighting during the GRPO process to achieve a Pareto-optimal trade-off between LLM-based semantic understanding and recommendation-oriented collaborative signals.

\subsubsection{LLM Semantic Reward}
To evaluate the semantic alignment between the LLM-generated reasoning output and the ground-truth target item, we propose an \textbf{LLM semantic reward} $r_s$ that operates in the textual semantic space. Specifically, we employ the Qwen3-Embedding-0.6B \cite{zhang2025qwen3emb} model to encode both the LLM-predicted answer (extracted from the \texttt{<answer>...</answer>} tags) and the ground-truth item description into dense semantic embeddings. The semantic reward is then computed as the cosine similarity between these two embeddings, formally defined as:
\vspace{-1pt}
\begin{equation} 
r_s = \text{CosineSim}(\mathbf{e}_{\text{pred}}, \mathbf{e}_{\text{gt}}),
\end{equation}
where $\mathbf{e}_{\text{pred}} = \text{Qwen3-Emb}(\text{Answer}_{\text{LLM}})$ and $\mathbf{e}_{\text{gt}} = \text{Qwen3-Emb}(\text{Item}_{\text{gt}})$ denote the semantic embeddings of the predicted answer and the ground-truth item, respectively. The cosine similarity is computed as:
$
\text{CosineSim}(\mathbf{e}_{\text{pred}}, \mathbf{e}_{\text{gt}}) = \frac{\mathbf{e}_{\text{pred}} \cdot \mathbf{e}_{\text{gt}}}{\|\mathbf{e}_{\text{pred}}\| \|\mathbf{e}_{\text{gt}}\|}.
$

This reward signal captures the semantic coherence between the LLM's reasoning output and the actual user purchase, providing a language-grounded supervision signal that complements traditional recommendation metrics.

\subsubsection{Recommendation ID Collaborative Reward}
To capture the collaborative filtering signals in the numerical continuous space, we propose a \textbf{recommendation ID collaborative reward} $r_{\text{id}}$ derived from the online ranking model deployed in the advertising system. This ranking model leverages user features, item features, cross features, and historical interaction labels to provide accurate probabilistic predictions of user-item interaction behaviors. Specifically, to align the LLM's reasoning outputs with real-world user preference patterns, we adopt the \textbf{click-through and conversion rate (CTCVR)} as the recommendation ID collaborative reward:

\begin{equation}
r_{\text{id}} = \text{CTCVR}(u, i) = P(\text{click} \land \text{conversion} \mid u, i),
\end{equation}
where $u$ and $i$ denote the user and the predicted item, respectively. To balance optimization across different rewards, we apply \textit{min-max normalization} to the raw CTCVR values. The CTCVR metric reflects the joint probability of both click and conversion events, providing a comprehensive measure of user purchase intent.

\textbf{Offline Simulation Environment.} To address the latency issue inherent in online evaluation, we construct an offline simulation environment by sampling user-item pairs and their prediction scores from the production system. This simulated environment enables efficient reward evaluation during the RL training process without requiring real-time online deployment, significantly accelerating the policy optimization while maintaining high fidelity to real-world user behaviors.

\subsubsection{Pareto Optimality Policy Optimization}
\label{sec:2.3.3}
Inspired by recent advances in adaptive multi-domain optimization \cite{fan2024doge, lu2025lop}, we propose \textbf{POPO}, which adaptively re-weights heterogeneous reward signals during RL training so as to dynamically navigate the Pareto front between the LLM-side semantic reward $r_s$ and the recommendation-side collaborative reward $r_{id}$.

Concretely, let $\mathcal{K}=\{s,\, id\}$ denote the set of reward sources, $\mathcal{L}_k(\theta)$ the GRPO objective induced by reward $r_k$, and $w_k^{(t)}$ its weight at the $t$-th RL iteration. The \textbf{POPO update rule} is defined as

\begin{equation} \label{update 1}
\boldsymbol{w}^{(t)} \;=\; \frac{\tilde{\boldsymbol{w}}^{(t)}}{\sum_{k\in\mathcal{K}} \tilde{w}_k^{(t)}}, \quad  
{\tilde{\boldsymbol{w}}^{(t)}} \;=\; \tilde{\boldsymbol{w}}^{(t-1)} \odot \exp\!\left(\frac{\eta^{(t)}\, \boldsymbol{I}^{(t)}}{\mu}\right), 
\end{equation}
where $\eta^{(t)}$ is the learning rate, $\mu>0$ is a regularization factor, and $\odot$ denotes the Hadamard product. The gradient alignment indicator $\boldsymbol{I}^{(t)}$ is computed as:
\begin{equation}
    I_i^{(t)} \;=\; \Bigl\langle \nabla_\theta \mathcal{L}_i\!\bigl(\theta^{(t)}\bigr),\; \sum_{k\in\mathcal{K}} \nabla_\theta \mathcal{L}_k\!\bigl(\theta^{(t)}\bigr) \Bigr\rangle.
\end{equation}

Intuitively, $I_i^{(t)}$ measures the alignment between the gradient of the $i$-th reward and the aggregated gradient direction of all rewards. \textit{Whenever a reward objective exhibits a large gradient magnitude and its gradient is well-aligned with the other objectives, exploiting it yields cooperative progress across all rewards and is therefore assigned a higher weight. Conversely, if its gradient conflicts with the other objectives, its weight is automatically suppressed, preventing it from dragging the policy off the Pareto front.}

\textbf{Pareto-Optimality Guarantee}. The above update can be rigorously interpreted as a first-order solution to the following bi-level optimization problem \cite{lu2025lop}:

$$
\boldsymbol{w}^{*} \;\in\; \underset{\boldsymbol{w}\in \Delta^{|\mathcal{K}|}}{\arg\min}\; \sum_{i\in\mathcal{K}} \mathcal{L}_i\!\bigl(\theta^{*}(\boldsymbol{w})\bigr), \text{ s.t.} \quad \theta^{*}(\boldsymbol{w}) \;\in\; \underset{\theta}{\arg\min}\; \sum_{i\in\mathcal{K}} w_i\, \mathcal{L}_i(\theta),
$$
where $\Delta^{|\mathcal{K}|}=\{\boldsymbol{w}\in\mathbb{R}_{\ge 0}^{|\mathcal{K}|} \mid \sum_k w_k = 1\}$ is the probability simplex. The lower-level problem reduces to a standard weighted GRPO update over the policy parameters $\theta$, while the upper-level problem further optimizes the reward weights $\boldsymbol{w}$ on top of the best-response policy $\theta^{*}(\boldsymbol{w})$. The exponentiated-gradient update above can be shown to be a mirror-descent approximation \cite{beck2003mirror} of this bi-level problem, and any of its stationary points $(\boldsymbol{w}^{*}, \theta^{*}(\boldsymbol{w}^{*}))$ satisfies the Pareto-optimality condition: no policy $\theta'$ exists such that $\mathcal{L}_k(\theta') \le \mathcal{L}_k(\theta^{*})$ for all $k\in\mathcal{K}$ with at least one strict inequality. 

\textbf{POPO-light}: an lightweight approximate POPO. In industrial-scale RL training, explicitly computing the cross-domain gradient inner product $I_i^{(t)}$ at every step incurs non-negligible time and memory overhead. To strike a better trade-off between effectiveness and efficiency, we further introduce POPO-light, a simplified variant that approximates the per-reward optimization potential purely from rollout-level reward statistics:

\begin{equation} \label{update 2}
\boldsymbol{w}^{(t)} \;=\; \frac{\tilde{\boldsymbol{w}}^{(t)}}{\sum_{k\in\mathcal{K}} \tilde{w}_k^{(t)}},  \quad \tilde{w}_i^{(t)} \;=\; \frac{\sigma_i^{(t)}}{\bigl|\mu_i^{(t)}\bigr| + \varepsilon}, 
\end{equation}
where $\mu_i^{(t)}$ and $\sigma_i^{(t)}$ are respectively the within-group mean and standard deviation of reward $r_i$ at step $t$, and $\varepsilon$ is a small constant for numerical stability. This formulation is exactly the \textbf{coefficient of variation} of each reward: a reward with large within-group variance relative to its magnitude indicates that the current policy is still highly discriminable along that direction and therefore deserves a larger weight; once a reward saturates (vanishing variance), its weight is automatically attenuated. Since POPO-light depends solely on forward reward scalars and avoids any additional backward gradient inner-product computation, it can be seamlessly plugged into the GRPO training loop with essentially zero extra overhead, making it particularly suitable for web-scale industrial deployment.

\textbf{Training Objective.} Building upon the proposed POPO, we optimize the policy with GRPO under dynamically re-weighted rewards, so as to achieve a more favorable trade-off between LLM semantic understanding and user preference modeling in LLM4Rec. Concretely, the training loss is formulated as

\small
\begin{equation}
\begin{aligned}
 & \mathcal{L}_{\text{POPO}}(\theta) = {} \mathbb{E}\!\left[\, q \sim P(Q),\ \{o_i\}_{i=1}^{G} \sim \pi_{\theta_{\text{old}}}(O \mid q) \,\right] \\
& \frac{1}{G}\sum_{i=1}^{G}\frac{1}{|o_i|}\sum_{t=1}^{|o_i|} \Bigg\{ \min\!\left[ \rho_{i,t}(\theta)\,\hat{A}_{i,t},\ \mathrm{clip}\!\left(\rho_{i,t}(\theta),\, 1-\varepsilon,\, 1+\varepsilon\right) \hat{A}_{i,t} \right] - \beta\, \mathbb{D}_{\mathrm{KL}}\!\left[\pi_\theta \,\Vert\, \pi_{\text{ref}}\right] \Bigg\},
\end{aligned}    
\end{equation}
where $\rho_{i,t}(\theta) = \pi_\theta(o_{i,t}\!\mid\! q, o_{i,<t}) / \pi_{\theta_{\text{old}}}(o_{i,t}\!\mid\! q, o_{i,<t})$, and
$
\hat{A}_{i,t} \;=\; \frac{\tilde{r}_i - \operatorname{mean}\!\bigl(\{\tilde{r}_1, \tilde{r}_2, \dots, \tilde{r}_G\}\bigr)}{\operatorname{std}\!\bigl(\{\tilde{r}_1, \tilde{r}_2, \dots, \tilde{r}_G\}\bigr)}
$ is the group-normalized advantage. Here $G$ denotes the group size, i.e., the number of candidate responses roll out from $\pi_{\theta_{\text{old}}}$ for each query $q$, and $\varepsilon$, $\beta$ are the clipping ratio and the KL regularization coefficient, respectively. The scalar reward of each rollout is given by the POPO-weighted combination of the two heterogeneous reward sources,

\begin{equation}
    \tilde{r}_i \;=\; w_s^{(t)}\, r_s(o_i) \;+\; w_{\text{id}}^{(t)}\, r_{\text{id}}(o_i),
\end{equation}
where $w_s^{(t)}$ and $w_{\text{id}}^{(t)}$ are the adaptive weights at iteration $t$ from Eq. (\ref{update 1})/(\ref{update 2}), and $\pi_{\text{ref}}$ is the SFT-initialized reference policy obtained from Section~\ref{sec:2.2}.

\begin{table*}[t]
\centering
\caption{Offline performance comparison on the test dataset. \textbf{Bold} and \underline{underline} denote the best and the second-best results, respectively. \emph{Semantic Hit-Rate} measures the LLM's recommendation reasoning quality (multi-level category accuracy and item-title Hit-Rate), while \emph{Preference Hit-Rate} (CTCVR) reflects the alignment with online user preferences. \textcolor{mypink}{Pink text} indicates the relative improvement over the backbone model DeepSeek-R1-7B.}
\label{tab:offline_main}
\setlength{\tabcolsep}{4pt}
\renewcommand{\arraystretch}{1}
\resizebox{\linewidth}{!}{%
\begin{tabular}{cl|ccccc}
\toprule
\textbf{Metric} & \textbf{Method} & \textbf{DeepSeek-R1-7B} & \textbf{QwQ-32B} & \textbf{Taiji (ORFT)} & \textbf{Taiji (ORFT+POPO-light)} & \textbf{Taiji (ORFT+POPO)} \\
\midrule
\multirow{5}{*}{\shortstack{Semantic\\Hit-Rate}}
& Category\_L1\_ACC & 0.1560 & 0.1974 & 0.2012 & \underline{0.2347} \textcolor{mypink}{($\uparrow$ 50.45\%)} & \textbf{0.2433} \textcolor{mypink}{($\uparrow$ 55.96\%)} \\
& Category\_L2\_ACC & 0.0608 & 0.0767 & 0.0690 & \underline{0.0877} \textcolor{mypink}{($\uparrow$ 44.24\%)} & \textbf{0.0888} \textcolor{mypink}{($\uparrow$ 46.05\%)}\\
& Category\_L3\_ACC & 0.0251 & 0.0039 & 0.0245 & \underline{0.0307} \textcolor{mypink}{($\uparrow$ 22.31\%)} & \textbf{0.0347} \textcolor{mypink}{($\uparrow$ 38.25\%)}\\
& Title\_Hit-Rate@50 & 0.0496 & \underline{0.0563} & 0.0449 & 0.0558 \textcolor{mypink}{($\uparrow$ 12.50\%)} & \textbf{0.0567} \textcolor{mypink}{($\uparrow$ 14.31\%)} \\
& Title\_Hit-Rate@100 & 0.0646 & \underline{0.0733} & 0.0606 & \textbf{0.0762} \textcolor{mypink}{($\uparrow$ 17.96\%)} & 0.0720 \textcolor{mypink}{($\uparrow$ 11.46\%)} \\
\midrule
\multirow{2}{*}{\shortstack{Preference\\Hit-Rate}}
& \multirow{2}{*}{CTCVR}
& \multirow{2}{*}{0.003417}
& \multirow{2}{*}{0.003675}
& \multirow{2}{*}{0.003723}
& \multirow{2}{*}{\underline{0.003802} \textcolor{mypink}{($\uparrow$ 11.27\%)}}
& \multirow{2}{*}{\textbf{0.003816} \textcolor{mypink}{($\uparrow$ 11.68\%)}}\\
& & & & & & \\
\bottomrule
\end{tabular}%
}
\end{table*}

\subsection{Online Ranking: Intra-User Features \& Cross-User Sequences Enhancement}
\label{sec:2.4}
After the RL-based policy optimization, the fine-tuned LLM is deployed for near-online inference to enhance the online ad ranking system. As illustrated in Figure~\ref{g3}, this Stage 4 bridges the gap between LLM reasoning and traditional recommendation models through two complementary mechanisms: intra-user feature quantization and cross-user sequence retrieval.

\textbf{Intra-User Feature Quantization.} For each incoming user request, we extract the user profile and behavioral history from online logs and feed them into the post-trained LLM to generate reasoning outputs in the format \texttt{<think>CoT</think>; <answer>Item Info</answer>}. To convert the textual reasoning into numerical features compatible with the ranking model, we employ the Qwen-Embedding-0.6B \cite{zhang2025qwen3emb} to encode the \textit{CoT\&Item Info} into dense embeddings. These embeddings are then quantized using product quantization~\cite{lian2020pq} to produce sparse ID vectors that capture the LLM's personalized understanding of user preferences. 

\textbf{Cross-User Sequence Retrieval.} To leverage collaborative signals across users, we perform similarity-based retrieval in the embedding space. Specifically, for a target user $u$, we retrieve the top-$1$ most similar users based on cosine similarity between their LLM-generated embeddings. The recent-100 behavioral sequences of the similar user are then aggregated and fed as additional features to the ranking model, enabling the model to capture cross-user interaction patterns that complement intra-user reasoning.

\textbf{Integration with Ranking Model.} Both the quantized sparse features and the retrieved cross-user sequences are concatenated with traditional features and fed into the online ad ranking model. This hybrid feature representation allows the ranking model to benefit from both the deep semantic understanding of LLMs and the collaborative filtering signals from user interactions, leading to improved ranking performance in production.


\section{Experiments}

In this section, we conduct extensive offline/online experiments and
detailed ablation studies based Kuaishou’s Advertising services, to validate the effectiveness and scalability of our Taiji.

\subsection{Experimental Setup}
\subsubsection{Datasets} 
We sample 1.11 million user records from the production system, each containing user profile information and behavioral sequences from the past month. The dataset is partitioned as follows: 1 million samples are allocated for SFT (Stage 2), where we generate $k=3$ recommendation CoT samples per user using QwQ-32B as the teacher model. The PPL truncation threshold is set to $R=4.6$, determined as the median PPL value computed over a validation set of 2.3K samples. An additional 100K samples are randomly selected for RL (Stage 3), and the remaining 10K samples constitute the test dataset. The ground-truth labels primarily consist of item titles and third-level category information.



\subsubsection{Evaluation Metrics}
For offline evaluation, we measure LLM performance using accuracy (ACC), hit rate at top-50 and top-100 (Hit-Rate@50, Hit-Rate@100), and click-through and conversion rate (CTCVR). For online A/B testing, we adopt advertiser value (ADVV) \cite{longer} and platform Revenue \cite{qarm} as key business metrics to assess the impact on advertiser ROI and platform monetization, respectively.



\subsubsection{Implementation Details} 
We employ DeepSeek-R1-7B\footnote{https://huggingface.co/deepseek-ai/DeepSeek-R1-Distill-Qwen-7B} as the base model and conduct training on 3 nodes with 8*A800 GPUs each. During the SFT stage, we perform full-parameter fine-tuning for 1 epoch with a learning rate of $1 \times 10^{-7}$ and a per-GPU batch size of 32. In the RL stage, we train for 1 epoch with a learning rate of $2 \times 10^{-5}$, a per-GPU batch size of 16, a rollout group size of $G=4$, a maximum prompt length of 13,000 tokens, and a maximum completion length of 2,048 tokens. The reward weights are initialized as $w_s = 0.5$ and $w_{\text{id}} = 0.5$.


\subsection{Offline Experiments}
\noindent We compare Taiji against two strong base LLMs: DeepSeek-R1-7B \cite{guo2025deepseek-r1} (the un-tuned backbone of Taiji), and QwQ-32B \cite{qwq-32b-preview} (the teacher model used for CoT distillation). We evaluate two complementary aspects of recommendation quality on the held-out 10K test dataset: (i) \emph{Semantic Hit-Rate}, which reflects how accurately the LLM reasons about user preferences in the textual space (multi-level category accuracy and item-title Hit-Rate@$\{50,100\}$); and (ii) \emph{Preference Hit-Rate}, measured by the offline-simulated CTCVR, which reflects how well the LLM's outputs align with online user behavior signals. The results are summarized in Table~\ref{tab:offline_main}.

Several observations can be drawn: 
\begin{itemize}
    \item \textbf{ORFT effectively activates reasoning capabilities.} \textit{Taiji (ORFT)} demonstrates substantial improvements over DeepSeek-R1-7B on category-level semantic understanding (+28.97\% on Category\_L1, +13.49\% on Category\_L2) and user preference alignment (+8.96\% on CTCVR), validating the effectiveness of reverse-engineered distillation. However, we observe a performance trade-off on title-level hit-rates, attributed to the \textit{limited generalization of SFT} while ORFT only successfully learns coarse-grained user-item matching patterns from training samples. Additionally, QwQ-32B's anomalously low Category\_L3 accuracy reflects its tendency toward overly general predictions without domain-specific fine-tuning. These limitations motivate the necessity of RL-based optimization to balance semantic understanding with lexical precision.
    \item \textbf{POPO yields consistent gains over both the 32B teacher and the SFT-only variant.} Equipping ORFT with POPO substantially boosts almost metrics and surpasses QwQ-32B (\textit{e.g.}, +23.25\% on Category\_L1 accuracy, and +3.84\% on CTCVR) and DeepSeek-R1-7B (\textit{e.g.}, +14.31\% on Title\_Hit-Rate@50, and 11.68\% on CTCVR), confirming that dynamically balancing semantic and ID-collaborative rewards effectively activates fine-grained recommendation preference signals in a much smaller 7B model. 
    \item \textbf{POPO and POPO-light are complementary.} The full POPO achieves the best score on five out of six metrics, while the lightweight variant POPO-light, despite using only rollout-level reward statistics and incurring negligible overhead, attains the second-best score on most metrics and even outperforms POPO on Title\_Hit-Rate@100. This makes POPO-light an attractive deployment choice when GPU budget is tight, while POPO remains the preferred option when training cost is not the bottleneck.
\end{itemize}
\begin{table*}[t]
\centering
\caption{Online A/B test results of Advertising services. Taiji demonstrates significant improvements in both ADVV and Revenue across different user segments. The \textit{long-tail} setting focuses on users with sparse interaction histories.}
\label{tab:online_results}
\setlength{\tabcolsep}{5pt}
\renewcommand{\arraystretch}{1.15}
\begin{tabular}{llcccccc}
\toprule
\multirow{2}{*}{\textbf{Method}} & \multirow{2}{*}{\textbf{Setting}} & \multicolumn{2}{c}{\textbf{Intra-User Features Enhancement}} & \multicolumn{2}{c}{\textbf{Cross-User Sequences Enhancement}} & \multicolumn{2}{c}{\textbf{Overall}} \\
\cmidrule(lr){3-4} \cmidrule(lr){5-6} \cmidrule(lr){7-8}
& & \textbf{ADVV} & \textbf{Revenue} & \textbf{ADVV} & \textbf{Revenue} & \textbf{ADVV} & \textbf{Revenue} \\
\midrule
\multirow{2}{*}{\textbf{Taiji}} 
& all & \textcolor{mypink}{+1.06\%} & \textcolor{mypink}{+1.35\%} & \textcolor{mypink}{+1.77\%} & \textcolor{mypink}{+1.95\%} & \textcolor{mypink}{+2.83\%} & \textcolor{mypink}{+3.30\%} \\
& long-tail & \textcolor{mypink}{+2.78\%} & \textcolor{mypink}{+4.12\%} & \textcolor{mypink}{+2.48\%} & \textcolor{mypink}{+1.20\%} & \textcolor{mypink}{+5.26\%} & \textcolor{mypink}{+5.32\%} \\
\bottomrule
\end{tabular}
\end{table*}

\subsection{Ablation Study}
To further validate the critical roles of RUPR (Section~\ref{sec:2.1.2}) and POPO (Section~\ref{sec:2.3.3}) in the Taiji framework, we conduct comprehensive ablation studies, as shown in Table~\ref{tab:ablation_rupr} and Table~\ref{tab:ablation_popo}.

\subsubsection{Impact of Reverse-Engineered User Preference Reasoning (RUPR)}
\begin{table}[t]
\centering
\caption{Ablation study on the RUPR module. \textit{w/o RUPR} uses CoT and item predictions directly generated by QwQ-32B, while \textit{w/ RUPR} leverages ground-truth labels from online logs to guide reasoning. \textcolor{mypink}{Pink text} indicates the relative improvement over directed distillation.}
\label{tab:ablation_rupr}
\setlength{\tabcolsep}{2.8pt}
\renewcommand{\arraystretch}{1.1}
\begin{tabular}{llccc}
\toprule
\textbf{Metric Type} & \textbf{Metric} & \textbf{w/o RUPR} & \textbf{w/ RUPR} & \textbf{Improv.} \\
\midrule
\multirow{3}{*}{\shortstack{Format\\Accuracy}}
& Think\_Tag\_Presence\_Rate & 0.9613 & 0.9947 & \textcolor{mypink}{+3.47\%} \\
& Think\_Non\_Empty\_Rate & 0.9613 & 0.9921 & \textcolor{mypink}{+3.20\%} \\
& Answer\_Non\_Empty\_Rate & 0.9607 & 0.9892 & \textcolor{mypink}{+2.97\%} \\
\midrule
\multirow{5}{*}{\shortstack{Semantics\\Hit-Rate}}
& Category\_L1\_ACC & 0.1679 & 0.2012 & \textcolor{mypink}{+19.83\%} \\
& Category\_L2\_ACC & 0.0535 & 0.0690 & \textcolor{mypink}{+28.97\%} \\
& Category\_L3\_ACC & 0.0185 & 0.0245 & \textcolor{mypink}{+32.43\%} \\
& Title\_Hit-Rate@50 & 0.0405 & 0.0449 & \textcolor{mypink}{+10.86\%} \\
& Title\_Hit-Rate@100 & 0.0545 & 0.0606 & \textcolor{mypink}{+11.19\%} \\
\bottomrule
\end{tabular}
\end{table}

Table~\ref{tab:ablation_rupr} demonstrates two key findings. First, the format accuracy after the SFT stage is already high ($\sim$ 99\%), establishing a solid foundation for subsequent RL training. Second, \textit{Taiji (ORFT w/o RUPR)} uses CoT and item predictions directly generated by QwQ-32B from user profiles and behavior sequences, whereas \textit{Taiji (ORFT w/ RUPR)} leverages ground-truth labels sampled from online logs to guide the CoT generation, providing dual semantic guarantees. Consequently, RUPR significantly improves semantic hit-rates, particularly for fine-grained metrics such as Category\_L3\_Accuracy (+32.43\%).


\subsubsection{Impact of Pareto Optimality Policy Optimization (POPO)}
\begin{table}[t]
\centering
\caption{Ablation study on the POPO algorithm. GRPO uses fixed reward weights ($w_s = w_{\text{id}} = 0.5$), while POPO dynamically adjusts weights to achieve Pareto optimality. \textcolor{mypink}{Pink text} indicates the relative improvement over GRPO.}
\label{tab:ablation_popo}
\setlength{\tabcolsep}{2.5pt}
\renewcommand{\arraystretch}{1.2}
\small
\begin{tabular}{lccc}
\toprule
\textbf{Metric} & \textbf{ORFT+GRPO} & \textbf{ORFT+POPO-light} & \textbf{ORFT+POPO} \\
\midrule
Category\_L1\_ACC & 0.2180 & \underline{0.2347} \textcolor{mypink}{($\uparrow$ 7.66\%)} & \textbf{0.2433} \textcolor{mypink}{($\uparrow$ 11.61\%)} \\
Category\_L2\_ACC & 0.0806 & \underline{0.0877} \textcolor{mypink}{($\uparrow$ 8.81\%)} & \textbf{0.0888} \textcolor{mypink}{($\uparrow$ 10.17\%)} \\
Category\_L3\_ACC & 0.0269 & 0.0307 \textcolor{mypink}{($\uparrow$ 14.13\%)} & \textbf{0.0347} \textcolor{mypink}{($\uparrow$ 29.00\%)} \\
Title\_Hit-Rate@50 & 0.0512 & 0.0558 \textcolor{mypink}{($\uparrow$ 8.98\%)} & \textbf{0.0567} \textcolor{mypink}{($\uparrow$ 10.74\%)} \\
Title\_Hit-Rate@100 & 0.0698 & \textbf{0.0762} \textcolor{mypink}{($\uparrow$ 9.17\%)} & \underline{0.0720} \textcolor{mypink}{($\uparrow$ 3.15\%)} \\
\midrule
CTCVR & 0.003788 & \underline{0.003802} \textcolor{mypink}{($\uparrow$ 0.37\%)} & \textbf{0.003816} \textcolor{mypink}{($\uparrow$ 0.74\%)} \\
\bottomrule
\end{tabular}
\end{table}

Unlike POPO, which dynamically adjusts reward weights, GRPO \cite{shao2024GRPO} uses fixed weights ($w_s = w_{\text{id}} = 0.5$) for both the LLM semantic reward and the recommendation ID collaborative reward. As shown in Table~\ref{tab:ablation_popo}, \textit{Taiji (ORFT+POPO)} simultaneously \textbf{improves both} \textbf{Semantic Hit-Rate} and \textbf{Preference Hit-Rate} compared to \textit{Taiji (ORFT+GRPO)}, validating that \textbf{POPO indeed pushes the policy toward the Pareto front rather than trading one objective off for the other}—the central design goal of our framework.

\subsection{Online Performance}
\label{sec:online_performance}

To validate the real-world effectiveness of Taiji, we conduct large-scale A/B tests on Kuaishou's advertising recommendation platform. We allocate 10\% of traffic to the baseline and 10\% to Taiji, running the experiment for one week. As shown in Table~\ref{tab:online_results}, Taiji achieves significant improvements in both ADVV (Advertiser Value) and Revenue (Platform Revenue). Notably, the gains are more pronounced for long-tail users with sparse interaction histories, where ADVV and Revenue increase by +5.26\% and +5.32\%, respectively. These results demonstrate that Taiji's reasoning-enhanced recommendation effectively bridges the semantic gap between user intent and item attributes, particularly benefiting users with limited behavioral data.

\section{Related Works}
\textbf{LLM-as-Enhancer via Reinforcement Alignment.} 
To bridge the semantic gap between LLM outputs and recommendation
objectives, recent work explores RL-based alignment for the
LLM-as-Enhancer paradigm. RecLM~\cite{jiang2025reclm} applies PPO with
an LLM-side semantic reward to refine generated user profiles, while
DEEPER~\cite{chen2025deeper} drives a tri-objective offline RL+DPO
loop with behavior-prediction discrepancy as the recommendation-side
signal. Rec-R1~\cite{lin2025Rec-R1} closes the loop with a black-box
recommender via GRPO, jointly optimizing semantic and NDCG/Recall
rewards, and LangPTune~\cite{gao2025langptune} jointly trains a
profile encoder and recommendation decoder through RLSO with
contrastive learning. RecGPT-V2~\cite{yi2025recgpt} further scales
this paradigm with a hierarchical multi-agent system and constrained
RL. However, these methods either rely on a single reward source or
combine heterogeneous rewards with \emph{static}, hand-crafted
weights, failing to characterize the non-convex Pareto front between
LLM world knowledge and online user preferences.

\section{Conclusions}
In this paper, we presented \textbf{Taiji}, an industrial-scale LLM-as-Enhancer framework that addresses two key limitations in the LLM4Rec post-training pipeline. In the SFT stage, we proposed Reverse-Engineered User Preference Reasoning (RUPR) together with Open-Ended Rejection sampling Fine-Tuning (ORFT) to distill and curate high-quality recommendation-specific CoT data. In the RL stage, we further introduced Pareto Optimal Policy Optimization (POPO), which adaptively re-weights the LLM semantic reward and the recommendation collaborative reward, achieving a theoretically grounded Pareto-optimal trade-off between LLMs' world knowledge and online user preferences. Extensive offline experiments and online A/B tests validate the effectiveness of Taiji, which has been fully deployed on Kuaishou's Advertising platform, stably serving over 400 million users.

\bibliographystyle{ACM-Reference-Format}
\bibliography{sample-base}


\begin{thebibliography}{30}


\ifx \showCODEN    \undefined \def \showCODEN     #1{\unskip}     \fi
\ifx \showISBNx    \undefined \def \showISBNx     #1{\unskip}     \fi
\ifx \showISBNxiii \undefined \def \showISBNxiii  #1{\unskip}     \fi
\ifx \showISSN     \undefined \def \showISSN      #1{\unskip}     \fi
\ifx \showLCCN     \undefined \def \showLCCN      #1{\unskip}     \fi
\ifx \shownote     \undefined \def \shownote      #1{#1}          \fi
\ifx \showarticletitle \undefined \def \showarticletitle #1{#1}   \fi
\ifx \showURL      \undefined \def \showURL       {\relax}        \fi
\providecommand\bibfield[2]{#2}
\providecommand\bibinfo[2]{#2}
\providecommand\natexlab[1]{#1}
\providecommand\showeprint[2][]{arXiv:#2}

\bibitem[Beck and Teboulle(2003)]%
        {beck2003mirror}
\bibfield{author}{\bibinfo{person}{Amir Beck} {and} \bibinfo{person}{Marc Teboulle}.} \bibinfo{year}{2003}\natexlab{}.
\newblock \showarticletitle{Mirror descent and nonlinear projected subgradient methods for convex optimization}.
\newblock \bibinfo{journal}{\emph{Operations Research Letters}} \bibinfo{volume}{31}, \bibinfo{number}{3} (\bibinfo{year}{2003}), \bibinfo{pages}{167--175}.
\newblock


\bibitem[Chai et~al\mbox{.}(2025)]%
        {longer}
\bibfield{author}{\bibinfo{person}{Zheng Chai}, \bibinfo{person}{Qin Ren}, \bibinfo{person}{Xijun Xiao}, \bibinfo{person}{Huizhi Yang}, \bibinfo{person}{Bo Han}, \bibinfo{person}{Sijun Zhang}, \bibinfo{person}{Di Chen}, \bibinfo{person}{Hui Lu}, \bibinfo{person}{Wenlin Zhao}, \bibinfo{person}{Lele Yu}, \bibinfo{person}{Xionghang Xie}, \bibinfo{person}{Shiru Ren}, \bibinfo{person}{Xiang Sun}, \bibinfo{person}{Yaocheng Tan}, \bibinfo{person}{Peng Xu}, \bibinfo{person}{Yuchao Zheng}, {and} \bibinfo{person}{Di Wu}.} \bibinfo{year}{2025}\natexlab{}.
\newblock \showarticletitle{LONGER: Scaling Up Long Sequence Modeling in Industrial Recommenders}. In \bibinfo{booktitle}{\emph{Proceedings of the Nineteenth ACM Conference on Recommender Systems}} \emph{(\bibinfo{series}{RecSys '25})}. \bibinfo{publisher}{Association for Computing Machinery}, \bibinfo{address}{New York, NY, USA}, \bibinfo{pages}{247–256}.
\newblock
\showISBNx{9798400713644}
\href{https://doi.org/10.1145/3705328.3748065}{doi:\nolinkurl{10.1145/3705328.3748065}}


\bibitem[Chen et~al\mbox{.}(2025)]%
        {chen2025deeper}
\bibfield{author}{\bibinfo{person}{Aili Chen}, \bibinfo{person}{Chengyu Du}, \bibinfo{person}{Jiangjie Chen}, \bibinfo{person}{Jinghan Xu}, \bibinfo{person}{Yikai Zhang}, \bibinfo{person}{Siyu Yuan}, \bibinfo{person}{Zulong Chen}, \bibinfo{person}{Liangyue Li}, {and} \bibinfo{person}{Yanghua Xiao}.} \bibinfo{year}{2025}\natexlab{}.
\newblock \showarticletitle{Deeper insight into your user: Directed persona refinement for dynamic persona modeling}. In \bibinfo{booktitle}{\emph{Proceedings of the 63rd Annual Meeting of the Association for Computational Linguistics (Volume 1: Long Papers)}}. \bibinfo{pages}{24157--24180}.
\newblock


\bibitem[Deng et~al\mbox{.}(2025)]%
        {deng2025onerec}
\bibfield{author}{\bibinfo{person}{Jiaxin Deng}, \bibinfo{person}{Shiyao Wang}, \bibinfo{person}{Kuo Cai}, \bibinfo{person}{Lejian Ren}, \bibinfo{person}{Qigen Hu}, \bibinfo{person}{Weifeng Ding}, \bibinfo{person}{Qiang Luo}, {and} \bibinfo{person}{Guorui Zhou}.} \bibinfo{year}{2025}\natexlab{}.
\newblock \showarticletitle{Onerec: Unifying retrieve and rank with generative recommender and iterative preference alignment}.
\newblock \bibinfo{journal}{\emph{arXiv preprint arXiv:2502.18965}} (\bibinfo{year}{2025}).
\newblock


\bibitem[Fan et~al\mbox{.}(2024)]%
        {fan2024doge}
\bibfield{author}{\bibinfo{person}{Simin Fan}, \bibinfo{person}{Matteo Pagliardini}, {and} \bibinfo{person}{Martin Jaggi}.} \bibinfo{year}{2024}\natexlab{}.
\newblock \showarticletitle{DOGE: Domain Reweighting with Generalization Estimation}. In \bibinfo{booktitle}{\emph{International Conference on Machine Learning}}. PMLR, \bibinfo{pages}{12895--12915}.
\newblock


\bibitem[Fleshman and Van~Durme(2025)]%
        {RE-AdaptIR}
\bibfield{author}{\bibinfo{person}{William Fleshman} {and} \bibinfo{person}{Benjamin Van~Durme}.} \bibinfo{year}{2025}\natexlab{}.
\newblock \showarticletitle{RE-AdaptIR: Improving Information Retrieval through Reverse Engineered Adaptation} \emph{(\bibinfo{series}{SIGIR '25})}. \bibinfo{publisher}{Association for Computing Machinery}, \bibinfo{address}{New York, NY, USA}, \bibinfo{pages}{2632–2636}.
\newblock
\showISBNx{9798400715921}
\href{https://doi.org/10.1145/3726302.3730240}{doi:\nolinkurl{10.1145/3726302.3730240}}


\bibitem[Gao et~al\mbox{.}(2025)]%
        {gao2025langptune}
\bibfield{author}{\bibinfo{person}{Zhaolin Gao}, \bibinfo{person}{Joyce Zhou}, \bibinfo{person}{Yijia Dai}, {and} \bibinfo{person}{Thorsten Joachims}.} \bibinfo{year}{2025}\natexlab{}.
\newblock \showarticletitle{LangPTune: Optimizing Language-based User Profiles for Recommendation}. In \bibinfo{booktitle}{\emph{Proceedings of the 34th ACM International Conference on Information and Knowledge Management}}. \bibinfo{pages}{707--717}.
\newblock


\bibitem[Gu et~al\mbox{.}(2025)]%
        {gu2025r4ec}
\bibfield{author}{\bibinfo{person}{Hao Gu}, \bibinfo{person}{Rui Zhong}, \bibinfo{person}{Yu Xia}, \bibinfo{person}{Wei Yang}, \bibinfo{person}{Chi Lu}, \bibinfo{person}{Peng Jiang}, {and} \bibinfo{person}{Kun Gai}.} \bibinfo{year}{2025}\natexlab{}.
\newblock \showarticletitle{R 4ec: A reasoning, reflection, and refinement framework for recommendation systems}. In \bibinfo{booktitle}{\emph{Proceedings of the Nineteenth ACM Conference on Recommender Systems}}. \bibinfo{pages}{411--421}.
\newblock


\bibitem[Guo et~al\mbox{.}(2025)]%
        {guo2025deepseek-r1}
\bibfield{author}{\bibinfo{person}{Daya Guo}, \bibinfo{person}{Dejian Yang}, \bibinfo{person}{Haowei Zhang}, \bibinfo{person}{Junxiao Song}, \bibinfo{person}{Peiyi Wang}, \bibinfo{person}{Qihao Zhu}, \bibinfo{person}{Runxin Xu}, \bibinfo{person}{Ruoyu Zhang}, \bibinfo{person}{Shirong Ma}, \bibinfo{person}{Xiao Bi}, {et~al\mbox{.}}} \bibinfo{year}{2025}\natexlab{}.
\newblock \showarticletitle{Deepseek-r1: Incentivizing reasoning capability in llms via reinforcement learning}.
\newblock \bibinfo{journal}{\emph{arXiv preprint arXiv:2501.12948}} (\bibinfo{year}{2025}).
\newblock


\bibitem[Jiang et~al\mbox{.}(2025)]%
        {jiang2025reclm}
\bibfield{author}{\bibinfo{person}{Yangqin Jiang}, \bibinfo{person}{Yuhao Yang}, \bibinfo{person}{Lianghao Xia}, \bibinfo{person}{Da Luo}, \bibinfo{person}{Kangyi Lin}, {and} \bibinfo{person}{Chao Huang}.} \bibinfo{year}{2025}\natexlab{}.
\newblock \showarticletitle{Reclm: Recommendation instruction tuning}. In \bibinfo{booktitle}{\emph{Proceedings of the 63rd Annual Meeting of the Association for Computational Linguistics (Volume 1: Long Papers)}}. \bibinfo{pages}{15443--15459}.
\newblock


\bibitem[Jiang et~al\mbox{.}(2026)]%
        {jiang2026tokenmixer}
\bibfield{author}{\bibinfo{person}{Yuchen Jiang}, \bibinfo{person}{Jie Zhu}, \bibinfo{person}{Xintian Han}, \bibinfo{person}{Hui Lu}, \bibinfo{person}{Kunmin Bai}, \bibinfo{person}{Mingyu Yang}, \bibinfo{person}{Shikang Wu}, \bibinfo{person}{Ruihao Zhang}, \bibinfo{person}{Wenlin Zhao}, \bibinfo{person}{Shipeng Bai}, {et~al\mbox{.}}} \bibinfo{year}{2026}\natexlab{}.
\newblock \showarticletitle{TokenMixer-Large: Scaling Up Large Ranking Models in Industrial Recommenders}.
\newblock \bibinfo{journal}{\emph{arXiv preprint arXiv:2602.06563}} (\bibinfo{year}{2026}).
\newblock


\bibitem[Li et~al\mbox{.}(2026)]%
        {li2026recgoat}
\bibfield{author}{\bibinfo{person}{Yuecheng Li}, \bibinfo{person}{Hengwei Ju}, \bibinfo{person}{Zeyu Song}, \bibinfo{person}{Wei Yang}, \bibinfo{person}{Chi Lu}, \bibinfo{person}{Peng Jiang}, {and} \bibinfo{person}{Kun Gai}.} \bibinfo{year}{2026}\natexlab{}.
\newblock \showarticletitle{RecGOAT: Graph Optimal Adaptive Transport for LLM-Enhanced Multimodal Recommendation with Dual Semantic Alignment}.
\newblock \bibinfo{journal}{\emph{arXiv preprint arXiv:2602.00682}} (\bibinfo{year}{2026}).
\newblock


\bibitem[Lian et~al\mbox{.}(2020)]%
        {lian2020pq}
\bibfield{author}{\bibinfo{person}{Defu Lian}, \bibinfo{person}{Xing Xie}, \bibinfo{person}{Enhong Chen}, {and} \bibinfo{person}{Hui Xiong}.} \bibinfo{year}{2020}\natexlab{}.
\newblock \showarticletitle{Product quantized collaborative filtering}.
\newblock \bibinfo{journal}{\emph{IEEE Transactions on Knowledge and Data Engineering}} \bibinfo{volume}{33}, \bibinfo{number}{9} (\bibinfo{year}{2020}), \bibinfo{pages}{3284--3296}.
\newblock


\bibitem[Lin et~al\mbox{.}(2025)]%
        {lin2025Rec-R1}
\bibfield{author}{\bibinfo{person}{Jiacheng Lin}, \bibinfo{person}{Tian Wang}, {and} \bibinfo{person}{Kun Qian}.} \bibinfo{year}{2025}\natexlab{}.
\newblock \showarticletitle{Rec-R1: Bridging Generative Large Language Models and User-Centric Recommendation Systems via Reinforcement Learning}.
\newblock \bibinfo{journal}{\emph{Transactions on Machine Learning Research}} (\bibinfo{year}{2025}).
\newblock
\showISSN{2835-8856}
\urldef\tempurl%
\url{https://openreview.net/forum?id=YBRU9MV2vE}
\showURL{%
\tempurl}


\bibitem[Lu et~al\mbox{.}(2025)]%
        {lu2025lop}
\bibfield{author}{\bibinfo{person}{Yining Lu}, \bibinfo{person}{Zilong Wang}, \bibinfo{person}{Shiyang Li}, \bibinfo{person}{Xin Liu}, \bibinfo{person}{Changlong Yu}, \bibinfo{person}{Qingyu Yin}, \bibinfo{person}{Zhan Shi}, \bibinfo{person}{Zixuan Zhang}, {and} \bibinfo{person}{Meng Jiang}.} \bibinfo{year}{2025}\natexlab{}.
\newblock \showarticletitle{Learning to optimize multi-objective alignment through dynamic reward weighting}.
\newblock \bibinfo{journal}{\emph{arXiv preprint arXiv:2509.11452}} (\bibinfo{year}{2025}).
\newblock


\bibitem[Luo et~al\mbox{.}(2025)]%
        {qarm}
\bibfield{author}{\bibinfo{person}{Xinchen Luo}, \bibinfo{person}{Jiangxia Cao}, \bibinfo{person}{Tianyu Sun}, \bibinfo{person}{Jinkai Yu}, \bibinfo{person}{Rui Huang}, \bibinfo{person}{Wei Yuan}, \bibinfo{person}{Hezheng Lin}, \bibinfo{person}{Yichen Zheng}, \bibinfo{person}{Shiyao Wang}, \bibinfo{person}{Qigen Hu}, \bibinfo{person}{Changqing Qiu}, \bibinfo{person}{Jiaqi Zhang}, \bibinfo{person}{Xu Zhang}, \bibinfo{person}{Zhiheng Yan}, \bibinfo{person}{Jingming Zhang}, \bibinfo{person}{Simin Zhang}, \bibinfo{person}{Mingxing Wen}, \bibinfo{person}{Zhaojie Liu}, {and} \bibinfo{person}{Guorui Zhou}.} \bibinfo{year}{2025}\natexlab{}.
\newblock \showarticletitle{QARM: Quantitative Alignment Multi-Modal Recommendation at Kuaishou} \emph{(\bibinfo{series}{CIKM '25})}. \bibinfo{publisher}{Association for Computing Machinery}, \bibinfo{address}{New York, NY, USA}, \bibinfo{pages}{5915–5922}.
\newblock
\showISBNx{9798400720406}
\href{https://doi.org/10.1145/3746252.3761502}{doi:\nolinkurl{10.1145/3746252.3761502}}


\bibitem[Rajput et~al\mbox{.}(2023)]%
        {rajput2023tiger}
\bibfield{author}{\bibinfo{person}{Shashank Rajput}, \bibinfo{person}{Nikhil Mehta}, \bibinfo{person}{Anima Singh}, \bibinfo{person}{Raghunandan Hulikal~Keshavan}, \bibinfo{person}{Trung Vu}, \bibinfo{person}{Lukasz Heldt}, \bibinfo{person}{Lichan Hong}, \bibinfo{person}{Yi Tay}, \bibinfo{person}{Vinh Tran}, \bibinfo{person}{Jonah Samost}, {et~al\mbox{.}}} \bibinfo{year}{2023}\natexlab{}.
\newblock \showarticletitle{Recommender systems with generative retrieval}.
\newblock \bibinfo{journal}{\emph{Advances in Neural Information Processing Systems}}  \bibinfo{volume}{36} (\bibinfo{year}{2023}), \bibinfo{pages}{10299--10315}.
\newblock


\bibitem[Shao et~al\mbox{.}(2024)]%
        {shao2024GRPO}
\bibfield{author}{\bibinfo{person}{Zhihong Shao}, \bibinfo{person}{Peiyi Wang}, \bibinfo{person}{Qihao Zhu}, \bibinfo{person}{Runxin Xu}, \bibinfo{person}{Junxiao Song}, \bibinfo{person}{Xiao Bi}, \bibinfo{person}{Haowei Zhang}, \bibinfo{person}{Mingchuan Zhang}, \bibinfo{person}{YK Li}, {et~al\mbox{.}}} \bibinfo{year}{2024}\natexlab{}.
\newblock \showarticletitle{Deepseekmath: Pushing the limits of mathematical reasoning in open language models}.
\newblock \bibinfo{journal}{\emph{arXiv preprint arXiv:2402.03300}} (\bibinfo{year}{2024}).
\newblock


\bibitem[Team(2024)]%
        {qwq-32b-preview}
\bibfield{author}{\bibinfo{person}{Qwen Team}.} \bibinfo{year}{2024}\natexlab{}.
\newblock \bibinfo{title}{QwQ: Reflect Deeply on the Boundaries of the Unknown}.
\newblock
\urldef\tempurl%
\url{https://qwenlm.github.io/blog/qwq-32b-preview/}
\showURL{%
\tempurl}


\bibitem[Wang et~al\mbox{.}(2025)]%
        {wang2025reer}
\bibfield{author}{\bibinfo{person}{Haozhe Wang}, \bibinfo{person}{Haoran Que}, \bibinfo{person}{Qixin Xu}, \bibinfo{person}{Minghao Liu}, \bibinfo{person}{Wangchunshu Zhou}, \bibinfo{person}{Jiazhan Feng}, \bibinfo{person}{Wanjun Zhong}, \bibinfo{person}{Wei Ye}, \bibinfo{person}{Tong Yang}, \bibinfo{person}{Wenhao Huang}, {et~al\mbox{.}}} \bibinfo{year}{2025}\natexlab{}.
\newblock \showarticletitle{Reverse-engineered reasoning for open-ended generation}.
\newblock \bibinfo{journal}{\emph{arXiv preprint arXiv:2509.06160}} (\bibinfo{year}{2025}).
\newblock


\bibitem[Xi et~al\mbox{.}(2024)]%
        {xi2024KAR}
\bibfield{author}{\bibinfo{person}{Yunjia Xi}, \bibinfo{person}{Weiwen Liu}, \bibinfo{person}{Jianghao Lin}, \bibinfo{person}{Xiaoling Cai}, \bibinfo{person}{Hong Zhu}, \bibinfo{person}{Jieming Zhu}, \bibinfo{person}{Bo Chen}, \bibinfo{person}{Ruiming Tang}, \bibinfo{person}{Weinan Zhang}, {and} \bibinfo{person}{Yong Yu}.} \bibinfo{year}{2024}\natexlab{}.
\newblock \showarticletitle{Towards open-world recommendation with knowledge augmentation from large language models}. In \bibinfo{booktitle}{\emph{Proceedings of the 18th ACM Conference on Recommender Systems}}. \bibinfo{pages}{12--22}.
\newblock


\bibitem[Xia et~al\mbox{.}(2025)]%
        {xia2025hit-lbm}
\bibfield{author}{\bibinfo{person}{Yu Xia}, \bibinfo{person}{Rui Zhong}, \bibinfo{person}{Hao Gu}, \bibinfo{person}{Wei Yang}, \bibinfo{person}{Chi Lu}, \bibinfo{person}{Peng Jiang}, {and} \bibinfo{person}{Kun Gai}.} \bibinfo{year}{2025}\natexlab{}.
\newblock \showarticletitle{Hierarchical tree search-based user lifelong behavior modeling on large language model}. In \bibinfo{booktitle}{\emph{Proceedings of the 48th International ACM SIGIR Conference on Research and Development in Information Retrieval}}. \bibinfo{pages}{1758--1767}.
\newblock


\bibitem[Xia et~al\mbox{.}(2026)]%
        {xia2025trackrec}
\bibfield{author}{\bibinfo{person}{Yu Xia}, \bibinfo{person}{Rui Zhong}, \bibinfo{person}{Zeyu Song}, \bibinfo{person}{Wei Yang}, \bibinfo{person}{Junchen Wan}, \bibinfo{person}{Qingpeng Cai}, \bibinfo{person}{Chi Lu}, {and} \bibinfo{person}{Peng Jiang}.} \bibinfo{year}{2026}\natexlab{}.
\newblock \showarticletitle{Trackrec: Iterative alternating feedback with chain-of-thought via preference alignment for recommendation}.
\newblock  (\bibinfo{year}{2026}).
\newblock


\bibitem[Yi et~al\mbox{.}(2025)]%
        {yi2025recgpt}
\bibfield{author}{\bibinfo{person}{Chao Yi}, \bibinfo{person}{Dian Chen}, \bibinfo{person}{Gaoyang Guo}, \bibinfo{person}{Jiakai Tang}, \bibinfo{person}{Jian Wu}, \bibinfo{person}{Jing Yu}, \bibinfo{person}{Mao Zhang}, \bibinfo{person}{Wen Chen}, \bibinfo{person}{Wenjun Yang}, \bibinfo{person}{Yujie Luo}, {et~al\mbox{.}}} \bibinfo{year}{2025}\natexlab{}.
\newblock \showarticletitle{RecGPT-V2 Technical Report}.
\newblock \bibinfo{journal}{\emph{arXiv preprint arXiv:2512.14503}} (\bibinfo{year}{2025}).
\newblock


\bibitem[Zhai et~al\mbox{.}(2024)]%
        {zhai2024hstu}
\bibfield{author}{\bibinfo{person}{Jiaqi Zhai}, \bibinfo{person}{Lucy Liao}, \bibinfo{person}{Xing Liu}, \bibinfo{person}{Yueming Wang}, \bibinfo{person}{Rui Li}, \bibinfo{person}{Xuan Cao}, \bibinfo{person}{Leon Gao}, \bibinfo{person}{Zhaojie Gong}, \bibinfo{person}{Fangda Gu}, \bibinfo{person}{Jiayuan He}, {et~al\mbox{.}}} \bibinfo{year}{2024}\natexlab{}.
\newblock \showarticletitle{Actions speak louder than words: trillion-parameter sequential transducers for generative recommendations}. In \bibinfo{booktitle}{\emph{Proceedings of the 41st International Conference on Machine Learning}}. \bibinfo{pages}{58484--58509}.
\newblock


\bibitem[Zhang et~al\mbox{.}(2025a)]%
        {zhang2025gpr}
\bibfield{author}{\bibinfo{person}{Jun Zhang}, \bibinfo{person}{Yi Li}, \bibinfo{person}{Yue Liu}, \bibinfo{person}{Changping Wang}, \bibinfo{person}{Yuan Wang}, \bibinfo{person}{Yuling Xiong}, \bibinfo{person}{Xun Liu}, \bibinfo{person}{Haiyang Wu}, \bibinfo{person}{Qian Li}, \bibinfo{person}{Enming Zhang}, {et~al\mbox{.}}} \bibinfo{year}{2025}\natexlab{a}.
\newblock \showarticletitle{GPR: Towards a Generative Pre-trained One-Model Paradigm for Large-Scale Advertising Recommendation}.
\newblock \bibinfo{journal}{\emph{arXiv preprint arXiv:2511.10138}} (\bibinfo{year}{2025}).
\newblock


\bibitem[Zhang et~al\mbox{.}(2026b)]%
        {zhang2026onemall}
\bibfield{author}{\bibinfo{person}{Kun Zhang}, \bibinfo{person}{Jingming Zhang}, \bibinfo{person}{Wei Cheng}, \bibinfo{person}{Yansong Cheng}, \bibinfo{person}{Jiaqi Zhang}, \bibinfo{person}{Hao Lu}, \bibinfo{person}{Xu Zhang}, \bibinfo{person}{Haixiang Gan}, \bibinfo{person}{Jiangxia Cao}, \bibinfo{person}{Tenglong Wang}, {et~al\mbox{.}}} \bibinfo{year}{2026}\natexlab{b}.
\newblock \showarticletitle{OneMall: One Model, More Scenarios--End-to-End Generative Recommender Family at Kuaishou E-Commerce}.
\newblock \bibinfo{journal}{\emph{arXiv preprint arXiv:2601.21770}} (\bibinfo{year}{2026}).
\newblock


\bibitem[Zhang et~al\mbox{.}(2025b)]%
        {zhang2025qwen3emb}
\bibfield{author}{\bibinfo{person}{Yanzhao Zhang}, \bibinfo{person}{Mingxin Li}, \bibinfo{person}{Dingkun Long}, \bibinfo{person}{Xin Zhang}, \bibinfo{person}{Huan Lin}, \bibinfo{person}{Baosong Yang}, \bibinfo{person}{Pengjun Xie}, \bibinfo{person}{An Yang}, \bibinfo{person}{Dayiheng Liu}, \bibinfo{person}{Junyang Lin}, {et~al\mbox{.}}} \bibinfo{year}{2025}\natexlab{b}.
\newblock \showarticletitle{Qwen3 embedding: Advancing text embedding and reranking through foundation models}.
\newblock \bibinfo{journal}{\emph{arXiv preprint arXiv:2506.05176}} (\bibinfo{year}{2025}).
\newblock


\bibitem[Zhang et~al\mbox{.}(2026a)]%
        {zhang2026onetrans}
\bibfield{author}{\bibinfo{person}{Zhaoqi Zhang}, \bibinfo{person}{Haolei Pei}, \bibinfo{person}{Jun Guo}, \bibinfo{person}{Tianyu Wang}, \bibinfo{person}{Yufei Feng}, \bibinfo{person}{Hui Sun}, \bibinfo{person}{Shaowei Liu}, {and} \bibinfo{person}{Aixin Sun}.} \bibinfo{year}{2026}\natexlab{a}.
\newblock \showarticletitle{Onetrans: Unified feature interaction and sequence modeling with one transformer in industrial recommender}. In \bibinfo{booktitle}{\emph{Proceedings of the ACM Web Conference 2026}}. \bibinfo{pages}{8162--8170}.
\newblock


\bibitem[Zhu et~al\mbox{.}(2025)]%
        {zhu2025rankmixer}
\bibfield{author}{\bibinfo{person}{Jie Zhu}, \bibinfo{person}{Zhifang Fan}, \bibinfo{person}{Xiaoxie Zhu}, \bibinfo{person}{Yuchen Jiang}, \bibinfo{person}{Hangyu Wang}, \bibinfo{person}{Xintian Han}, \bibinfo{person}{Haoran Ding}, \bibinfo{person}{Xinmin Wang}, \bibinfo{person}{Wenlin Zhao}, \bibinfo{person}{Zhen Gong}, {et~al\mbox{.}}} \bibinfo{year}{2025}\natexlab{}.
\newblock \showarticletitle{Rankmixer: Scaling up ranking models in industrial recommenders}. In \bibinfo{booktitle}{\emph{Proceedings of the 34th ACM International Conference on Information and Knowledge Management}}. \bibinfo{pages}{6309--6316}.
\newblock


\end{thebibliography}










\end{document}